\documentclass[twocolumn]{aastex631}

\newcommand{\ts}{t_{\rm stop}}
\newcommand{\cs}{c_{\rm s}}
\newcommand{\cd}{c_{\rm d}}
\newcommand{\sd}{\Sigma_{\rm d}}

\newcommand{\Ml}{M_{\rm L}}
\newcommand{\St}{{\rm St}}
\newcommand{\tml}{{\tilde M}_{\rm L}}
\newcommand{\tw}{{\tilde W}}
\newcommand{\tk}{{\tilde k}}

\usepackage{comment}
\usepackage{color}
 \usepackage{multirow}

\shorttitle{Ring GI in Protoplanetary Disks}
\shortauthors{Takahashi, Kokubo and Inutsuka}

\graphicspath{{./}{figures/}}

\begin{document}

\title{Planetesimal formation by the gravitational instability of dust ring structures}

\author{Sanemichi Z. Takahashi}
\affiliation{National Astronomical Observatory of Japan, 2-21-1 Osawa, Mitaka, Tokyo 181-8588, Japan}
\email{sanemichi.takahashi@nao.ac.jp}
\author{Eiichiro Kokubo}
\affiliation{National Astronomical Observatory of Japan, 2-21-1 Osawa, Mitaka, Tokyo 181-8588, Japan}
\author{Shu-ichiro Inutsuka}
\affiliation{Department of Physics, Nagoya University, Furo-cho, Chikusa-ku, Nagoya, 464-8602 Aichi, Japan}

\begin{abstract}

We investigate the gravitational instability (GI) of dust-ring structures and the formation of planetesimals by their gravitational collapse. The normalized dispersion relation of a self-gravitating ring structure includes two parameters that are related to its width and line mass (the mass per unit length). We survey these parameters and calculate the growth rate and wavenumber. Additionally, we investigate the planetesimal formation by growth of the GI of the ring that is formed by the growth of the secular GI of the protoplanetary disk. We adopt a massive, dust rich disk as a disk model. We find the range of radii for the fragmentation by the ring GI as a function of the width of the ring. The inner-most radius for the ring GI is smaller for the smaller ring width. We also determine the range of the initial planetesimal mass resulting from the fragmentation of the ring GI.  Our results indicate that the planetesimal mass can be as large as $10^{28}$ g at its birth after the fragmentation. It can be as low as about $10^{25}$ g if the ring width is 0.1\% of the ring radius and the lower limit increases with the ring width.  Furthermore, we obtain approximate formulas for the upper and lower limits of the planetesimal mass. We predict that the planetesimals formed by the ring GI have prograde rotations because of the Coriolis force acting on the contracting dust. This is consistent with the fact that many trans-Neptunian binaries exhibit prograde rotation.

\end{abstract}


\section{INTRODUCTION}

Protoplanetary disks are considered to be the planet formation sites. Recent high spatial resolution observations of protoplanetary disks have revealed that they contain a variety of detailed structures. 
In particular, observations with Atacama Large Millimeter/submillimeter Array (ALMA) reveal that many disks have ring-gap structures 
\cite[e.g.][]{2015ApJ...808L...3A,2016PhRvL.117y1101I,2016ApJ...820L..40A,2016ApJ...829L..35T,2017A&A...600A..72F,2017ApJ...840L..12S,2018ApJ...869L..41A}.
However, the origins of these structures have not been clarified.
The formation mechanisms proposed so far for the ring-gap structures include the formation of gap structures by planets \cite[e.g.][]{2015ApJ...806L..15K,2017PASJ...69...97K,2015ApJ...809...93D,2017ApJ...843..127D,2016ApJ...818...76J}, the evolution of dust due to sintering \cite[]{2016ApJ...821...82O}, secular gravitational instability (GI) \cite[]{2014ApJ...794...55T,2016AJ....152..184T,2018PASJ...70....3T,2019ApJ...881...53T,2020ApJ...900..182T}, magnetorotational instability (MRI) \cite[e.g.][]{2011ApJ...736...85U,2015A&A...574A..68F}, and disk wind \cite[e.g.][]{2018ApJ...865..102T} have been proposed. 

Although the ring structure formation mechanism is still under debate, this structure is important for the planet formation process. 
If a ring-gap structure is formed by a planet, the ring structure around a young protostar limits the planet formation timescale, since the planet should be formed within the time of the age of the protostar.
Meanwhile, if these ring structures are formed by the concentration of dust through a mechanism that does not require the involvement of planets, one expects planetesimals to be formed in the ring. 
Because ring structures are observed in several disks, the planetesimal formation in a ring structure may be important for the planet formation scenario. 

In this study, we consider a planetesimal formation scenario due to the gravitational collapse of a dust ring structure, which has not been considered so far. 
Secular GI, which is one of the mechanisms for forming ring structures, can form a self-gravitating dust ring. 
A one-dimensional simulation of the nonlinear growth of the secular GI  \cite[]{2020ApJ...900..182T} shows that the mass of a dust ring can be $\sim$ 10 $M_\oplus$ and the surface density of dust can be  more than 10 times than its initial value. 
Such a dense, massive ring can collapse because of GI.
When dust is concentrated in a ring, the collisional growth of the dust is promoted, resulting in  a decrease of friction between the dust and gas around the ring. 
Moreover, if the surface density of dust increases during ring formation, and becomes larger than the surface density of gas, the effect of friction between dust and gas becomes small. 
Therefore, after the massive ring structure formation via secular GI, the dust structure evolves owing to the GI of the dust component and not the secular GI in which dust-gas friction is important.
In this study, we investigate the process in which a ring structure formed by a secular GI  subsequently collapses because of the ring structure's GI (hereafter, referred to as ring GI). 

In this paper, we investigate a gravitational collapse where a dust ring structure collapses in the azimuthal direction along the ring. 
By the linear stability analysis, we obtain the dependence of the dispersion relation on the line mass (mass per unit length) of the ring and width.
Using this instability, we investigate a scenario in which a ring structure formed by the secular GI of the disk fragments by the ring GI.
We estimate the line mass of the ring from the most unstable wavelength of the secular GI. 
We perform a linear analysis of the GI of the ring structure to estimate the masses of the planetesimals formed by the ring's gravitational collapse.
For simplicity, we call the fragments formed by the dust ring's gravitational collapse as ``planetesimals,'' although their masses can be as large as the mass of a planet.
We estimate the masses of the planetesimals using the most unstable wavelength of the ring GI and the largest wavelength at which the instability can grow (the largest unstable wavelength). 
The mode with the most unstable wavelength grows fastest and collapses first. 
Subsequently, we expect an increase in longer wavelength perturbations and the masses of the corresponding planetesimals.
Consequently, the planetesimal masses grow to the masses estimated from the largest unstable wavelength. 
Therefore, in this study, we estimate a lower limit of planetesimal mass from the most unstable wavelength and an upper limit from the largest unstable wavelength.

This paper is organized as follows.
In Section \ref{disp}, we present the dispersion relation of a self-gravitating ring structure and investigate its dependence on the line mass and width of the ring.
In Section \ref{result}, we estimate the masses of the planetesimals formed by the ring GI in a ring structure produced by the secular GI of the disk. We also determine the dependence of the mass on the parameters such as the initial conditions of the disk and the width of the ring.
Sections \ref{discussion} and \ref{summary} provide some discussion and a summary, respectively.

\section{Dispersion relation of a self-gravitating ring structure}
\label{disp}

\subsection{Basic Equations and Dispersion Relation}

Here, we investigate the GI of a dust ring structure.
Such instabilities have been studied previously for starbursts in galaxies \cite[]{1994ApJ...425L..73E} and the fragmentation process of the spiral arm of a self-gravitating disk \cite[]{2016MNRAS.458.3597T}. 
The gravitational potential terms are different for these studies.
For the dispersion relation of these studies, the gravitational potential is estimated at the core curve of the ring in \cite[]{1994ApJ...425L..73E}, and is estimated considering the ring as an infinitely thin structure with a finite width in \cite[]{2016MNRAS.458.3597T}.
There are also the related previous studies \cite{1988MNRAS.231...97G} \cite{1997ApJS..108..471A}, and \cite{1986A&A...161..403W}.
The former two has investigated the stability of the self-gravitating incompressible ring. 
The latter one has investigated the stability of the self-gravitating ring by dividing the ring into discrete particles and solving their motion.
In this study, we use the equations given by \cite[]{2016MNRAS.458.3597T} adding the effect of thickness to the gravitational potential to investigate the GI of the ring structure of the dust which is spread vertically by the turbulence of the gas in protoplanetary disks.
Using these equations, we simplify the motion of the radial direction as explained below. This treatment corresponds to the omission of the radial oscillation modes of the ring, which obtained in the previous studies \cite{1988MNRAS.231...97G} \cite{1997ApJS..108..471A} and \cite{1986A&A...161..403W}, but these modes are stable and not responsible for the gravitational fragmentation of the ring.

We use the local approximation and employ shearing coordinates to derive the dispersion relation, where the radial and azimuthal direction is $x$ and $y$, respectively.
The basic equations used here is similar to the equations for the self-gravitating infinitesimally thin filament (equation of continuity, equation of motion in the $y$ direction, Poisson equation), whereas the equation of motion in the $x$ direction is added to take into account the effect of rotation, which contributes to stabilization in the case of a ring structure.
We assume that the ring is thin and consider the ring width and thickness only to evaluate the gravitational potential.
The basic equations are given as follows:
\begin{equation}
 -i \omega \delta \Ml+ ik_y\Ml\delta v_y = 0,
\label{eq:eoc}
\end{equation}
\begin{equation}
 -i\omega\delta v_x = 
2\Omega\delta v_y,
\label{eq:eom_x}
\end{equation}
\begin{equation}
 -i\omega \delta v_y = -\frac{1}{2}\Omega\delta v_x
-ik_y\frac{\cd^2}{\Ml}\delta\Ml-i\frac{k_y}{1+k_yH_{\rm d}}\delta\Phi,
\label{eq:eom_y}
\end{equation}
\begin{equation}
 \delta \Phi = -\pi G \delta \Ml g(k_yW),
\label{eq:poi}
\end{equation}
\begin{equation}
 g(k_yW)=[K_0(k_yW)L_{-1}(k_yW)+K_1(k_yW)L_0(k_yW)],
\label{eq:eq_GP}
\end{equation}
where $\omega$ and $k_y$ are the perturbation frequency and wavenumber; $\Ml$ and $W$ are the line mass and half width of the ring; $\cd$ is the velocity dispersion of the dust; $\Omega$ is the Kepler frequency; $H_{\rm d}=\cd/\Omega$ is the scale height of the ring; $\delta \Ml,\ \delta v_x,\ \delta v_y,\ \delta \Phi$ are the perturbations of the line mass, the $x$- and $y$-components of the velocity, and the gravitational potential, $K_0$, and $K_1$ are the modified Bessel functions of the second kind and $L_0$ and $L_{-1}$ are the modified Struve functions, respectively.
The coefficient of the first term in the right-hand side of Equation (\ref{eq:eom_y}) is different from that in the corresponding equation in \cite{2016MNRAS.458.3597T} (Equation (18)) because in the previous paper we adopted the rigid rotation as the background rotation profile, as suggested by the numerical simulations, while the Keplerian rotation is adopted for simplicity in this work.
\footnote{
There was a typo in the sign of the second term of Equation (16) in \cite{2016MNRAS.458.3597T} that corresponds to Equation (\ref{eq:eoc}).
}
Equation (\ref{eq:poi}) gives the perturbed gravitational potential of the flattened ring with the half width $W$. To calculate the gravitational potential, we divide the flattened ring into the series of
thin filaments and evaluate the gravitational potential from the integration of the gravitational potential of the filaments.
The perturbed gravitational potential of the thin filament is 
\begin{equation}
d(\delta \Phi) = -2GK_0(k_yx')d(\delta M_{\rm L}),
\end{equation}
where $x'$, $d(\delta \Phi)$, and $d(\delta M_{\rm L})$ are the distance from the axis of the filament, and the perturbed gravitational potential and the perturbed line mass of the filament, respectively \cite[]{1961hhs..book.....C}.
When we replace the perturbed line mass of the thin filament $d(\delta M_{\rm L})$ by $\delta \Sigma(x)dx$, the gravitational potential of the flattened ring is given by 
\begin{equation}
    \delta\Phi = \int d(\delta \Phi) = \int -2G\delta\Sigma(x)K_0(k_yx)dx.
\end{equation}
Here we assume that $\delta \Sigma (x) $ has a constant value $\delta M_{\rm L}/(2W)$ for $-W < x< W$ and 0 for $x<-W, x>W$ for simplicity, and obtain Equation (\ref{eq:poi}).
The third term on the right-hand side of Equation (\ref{eq:eom_y}) represents the gravitational force. The denominator $1+k_{\rm y} H_{\rm d}$ approximately gives the thickness effect, which is valid for $k_{\rm y} \lesssim H_{\rm d}^{-1}$ \cite[]{1970ApJ...161...87V,1984prin.conf..513S}.

We obtain the following dispersion relation from these equations (\ref{eq:eoc})--(\ref{eq:eq_GP}):
\begin{equation}
\omega^2 = \cd^2k_y^2 -\pi GM_{\rm L} g(k_yW)
\frac{k_y^2}{1+kH_{\rm d}} +\Omega^2.
\end{equation}
We normalize the dispersion relation using the Kepler frequency and the velocity dispersion,
\begin{equation}
 {\tilde \omega}^2 = {\tilde k}_y^2 -\pi \tml g({\tilde k}_y \tw)
\frac{{\tilde k}_y^2}{1+{\tilde k}_y} +1,
\label{eq:disp_norm}
\end{equation}
where
\begin{equation}
 {\tilde \omega} = \frac{\omega}{\Omega},
\end{equation}
\begin{equation}
 {\tilde k}_y = k_yH_{\rm d} =\frac{k_y\cd}{\Omega},
\end{equation}
\begin{equation}
 \tml=\frac{GM_{\rm L}}{\cd^2},
\end{equation}
\begin{equation}
 \tw = \frac{W}{H_{\rm d}}.
\end{equation}
Notably, the definitions of the normalized values $ {\tilde \omega}$ and $ {\tilde k}_y$ are different from those reported by \cite{2016MNRAS.458.3597T}; the new normalization makes the parameter dependence clearer than the previous one.
The parameters $\tml$ and $\tw$ represent the normalized line mass and ring width, respectively. 
The parameter $\tml$ is the same as the parameter $f$ in \cite{2016MNRAS.458.3597T}.

The proposed dispersion relation is similar to the well-known dispersion relation of a self-gravitating disk; only the term related to the gravitational force (the second term on the right-hand side) is different because the background is not a disk but a filament (a local part of the ring).

\begin{figure}[tb]
\includegraphics[width=8cm]{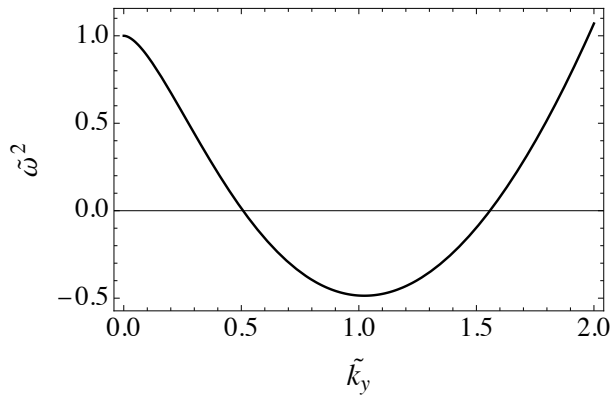}
\caption{Dispersion relation of a self-gravitating ring with $\tml=2$ and $\tw=1$. The vertical axis is the square of the normalized frequency, and the horizontal axis is the normalized wavenumber. The ring is unstable when ${\tilde \omega}^2<0$. }
\label{fig:ring_disp}
\end{figure}

In Figure \ref{fig:ring_disp}, we show the dispersion relation for a case with $\tml=2$ and $\tw=1$.
In this case, the function ${\tilde \omega} ^2({\tilde k})$ seems to be a parabolic function, except at small wavenumbers ($k_y\ll1$).

\subsection{Dependence on the ring width and the line mass}

\begin{figure}[tb]
\begin{minipage}{80mm}
\begin{center}
\plotone{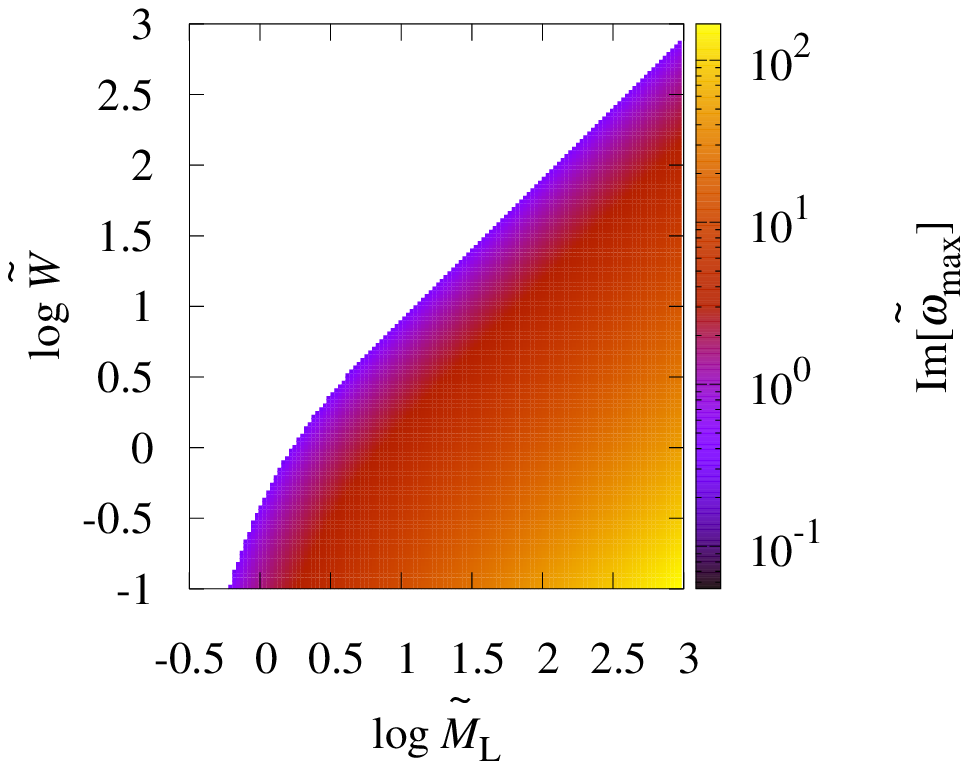}
\end{center}
\end{minipage}
\\
\begin{minipage}{80mm}
\begin{center}
\plotone{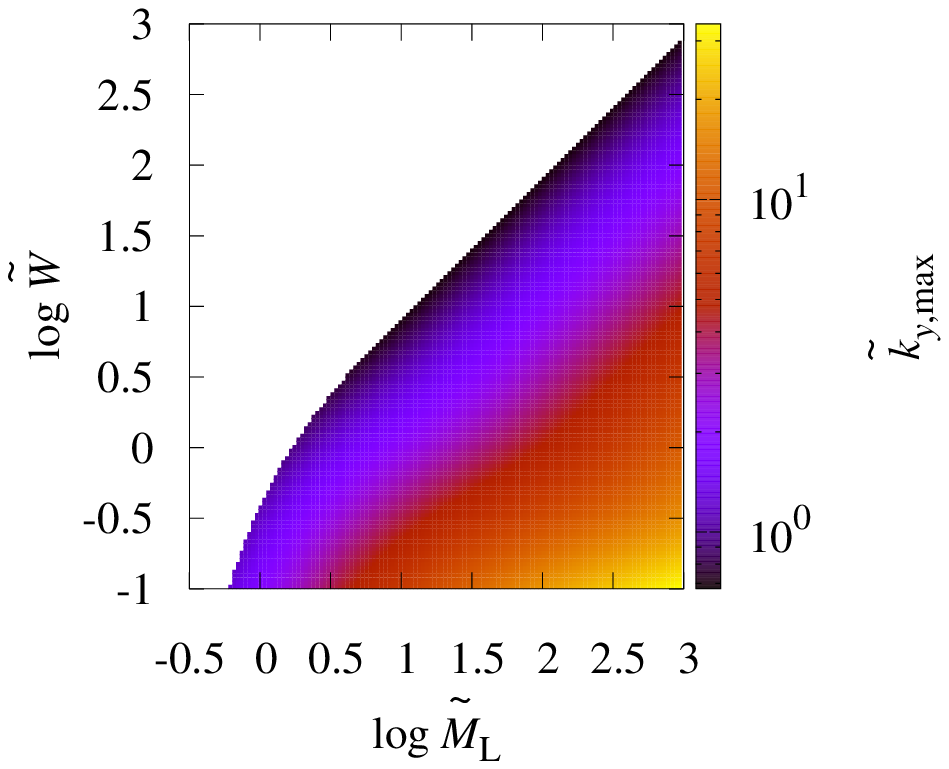}
\end{center}
\end{minipage}
\\
\begin{minipage}{80mm}
\begin{center}
\plotone{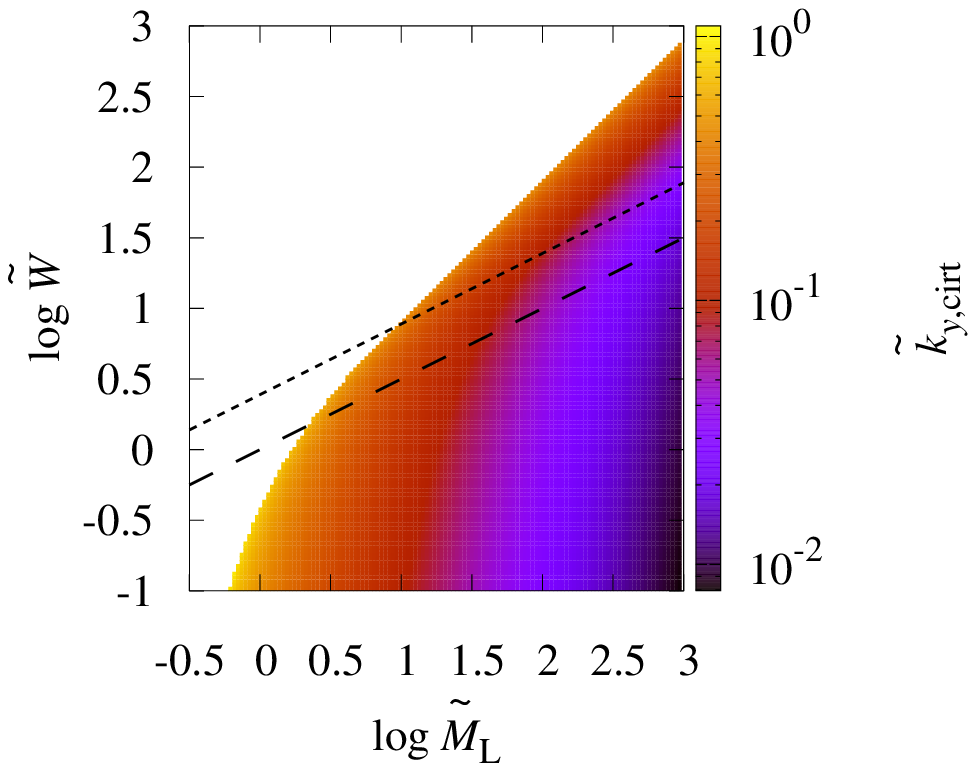}
\end{center}
\end{minipage}
\caption{Growth rate at the most unstable wavelength (top panel), the most unstable wavenumber (middle panel), and the smallest unstable wavenumber(bottom panel).
}
\label{fig:param_surv}
\end{figure}

The dispersion relation includes two parameters, $\tml$ and $\tw$, which are the normalized line mass and width of the ring, respectively.
We investigate the dependence  of the growth timescales and the most unstable and the largest unstable wavelength on these parameters.
In Figure \ref{fig:param_surv}, we show the growth rates at ${\tilde k}_{y,{\rm max}}$, ${\tilde k}_{y,{\rm max}}$, and ${\tilde k}_{y,{\rm crit}}$ of the instability for $10^{-0.5}\leq \tml\leq 10^3$ and $10^{-1}\leq \tw\leq10^3${\ , where ${\tilde k}_{y,{\rm max}}$ and ${\tilde k}_{y,{\rm crit}}$ are the most unstable and the smallest unstable normalized wavenumber, respectively}.
The ring is stable in the white regions of the panels.
The condition of the ring GI is similar to that for the disk structure's GI for large $\tml$ and $\tw$ values.
Please see the next subsection for an analytical description of the instability condition.
For small $\tml$ and $\tw$ values, the condition deviates from this relation owing to the effect of the ring structure.
For a ring with a small width and line mass, the most unstable wavelength is larger than the ring width.
For such a perturbation, the contribution of the mass to the self-gravity is small compared with that of the self-gravitating disk because the ring has a finite radial extent. 
Consequently, the effect of the self-gravity of the ring becomes smaller than that of the self-gravitating disk.
Thus, for small $\tml$, a smaller $\tw$ is required to cause ring instability than that required to produce disk instability .

\subsection{Analytical Formulas}
\label{app}

\begin{figure}[tb]
\includegraphics[width=8cm]{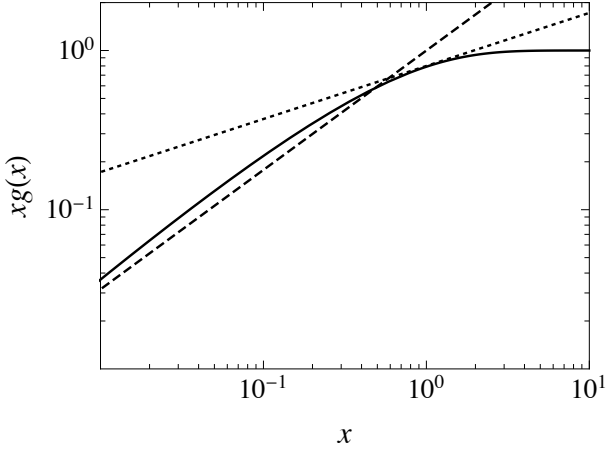}
\caption{Comparison of the function $xg(x)$ (solid line) and the approximate functions $(4/5)x^{1/3}$ (dotted line) and $x^{3/4}$ (dashed line). 
}
\label{fig:GP_fit}
\end{figure}

To understand the nature of the instability, it is useful to analytically derive the instability condition and the most unstable wavelength.
In this subsection, we estimate the instability condition and the most unstable wavelength by approximating Equation (\ref{eq:eq_GP}), which gives the wavenumber dependence of the gravitational potential.
Moreover, we estimate the smallest wavenumber for the instability, which we use in Section \ref{result} to estimate the upper limit of the planetesimal mass.
As shown in Figure \ref{fig:GP_fit} we find  the approximate dispersion relation for the wavelength larger than the ring width (${\tilde k}_y\tw\gg 1$), comparable to the ring width (${\tilde k}_y\tw\sim 1$), and smaller than the ring width (${\tilde k}_y\tw\ll1$);
for these wavelengths, the function $xg(x)$ can be roughly fitted by unity, $(4/5)x^{1/3}$, and $x^{3/4}$, respectively, where $x={\tilde k}_y{\tilde W}$. 
Here we give the fitting function for the intermediate case (${\tilde k}_y\tw\sim 1$) to make the fitting accurate in this case. This fitting is useful because the intermediate case appeared frequently in our previous study \cite[]{2016MNRAS.458.3597T}.

We summarize the normalized most unstable and smallest unstable wavenumber, and the instability condition in Table \ref{tab:app}.
The details are explained in the following subsections and Appendix \ref{deri}.

\begin{table*}
\caption{Normalized most unstable and smallest wavenumbers and  the instability condition}
\label{tab:app}
 \centering
  \begin{tabular}{c|c|c|c}
  \hline
   &${\tilde k}_{y,{\rm max}}$ & ${\tilde k}_{y,{\rm crit}}$ & Instability Condition\\
  \hline
    %%%%%%%%%%%%%%%%% small wavelength region %%%%%%%%%%%%%%%%%
  \multirow{2}{*}{${\tilde k}_y{\tilde W} \gtrsim 2 $ } 
  & $\frac{1}{2}+\frac{\pi \tml}{16\tw}$ ($\tk\sim 1$)                 %k_max  for k~1
  &\multirow{2}{*}{$ \frac{\pi \tml}{2 \tw}                            %k_crit
-\sqrt{\left(\frac{\pi \tml}{2\tilde{W}}\right)^2-1}$}
&\multirow{2}{*}{$\tml\simeq1.3\tw$}\\                               % condition 
  & $\left(\frac{\pi\tml}{2\tw} \right)^{1/3} $ ($\tk\gg 1 $)&&\\      %k_max  for k>>1
  %%%%%%%%%%%%%%%%%% intermediate wavelength region %%%%%%%%%%%%%%%%%
  $0.5 \lesssim {\tilde k}_y{\tilde W} \lesssim 2$  
  & $\left( \frac{2\pi\tml}{15\tw^{2/3}}\right)^{3/5}$                        %kmax
  &$ \left(\frac{5}{4 \pi \tml}\right)^{3/4}\tw^{1/2}$ %kcrit
  &$\tml\simeq1.7\tw^{2/3}$\\                                                 %condition
  %%%%%%%%%%%%%%%%%% large wavelength region %%%%%%%%%%%%%%%%%
  ${\tilde k}_y{\tilde W} \lesssim 0.5 $
  &$\left( \frac{3\pi\tml}{8\tw^{1/4}}\right)^{4/5}$                 %kmax
  & $ \frac{\tw^{1/7}}{(\pi \tml)^{4/7}}$     %kcrit
  &$\tml\simeq 1.2\tw^{1/4}$\\                                       %condition
  \hline
  \end{tabular}
\end{table*}

\subsubsection{Large wavenumber \texorpdfstring{$(\tk_y\tw \gtrsim 2)$}{kW>2}}

For large wavenumbers ($\tk_y\gg 1/\tw$), the term ${\tilde k}_y\tw g({\tilde k}_y\tw)$ approaches unity  asymptotically .
In this case, the dispersion relation is the same as that of a self-gravitating disk when we replace $\Ml/(2W)$ with the surface density $\Sigma$ or  replace $\pi \tml/\tw$ with $2/Q$, where $Q$ is the Toomre parameter.
The approximated dispersion relation is given as
\begin{equation}
 {\tilde \omega}^2={\tilde k}_y^2 -\frac{\pi \tml}{\tw}\frac{{\tilde k}_y}{1+{\tilde k}_y}+1.
 \label{eq:disp_short}
\end{equation}

As shown in Figure \ref{fig:param_surv}, the most unstable wavelength is approximately unity around the boundary of the unstable region, while it is much larger than unity in the region with large $\tml$ and small $\tw$ values. To obtain the approximate dispersion relation analytically, we derive the approximate most unstable wavelength for ${\tilde k}_y \sim 1 $ and  ${\tilde k}_y \gg 1$, separately. 
To drive the approximate formula for ${\tilde k}_{y,\rm max} \sim 1$ is required to obtain the approximate condition for the instability. 
Note that the case for ${\tilde k}_{y,\rm max} \ll 1$ does not appear in the parameter region shown in Figure \ref{fig:param_surv}.
The detailed derivations are given in Appendix \ref{deri}.

For ${\tilde k}_{y,{\rm max}} \sim 1$, the most unstable wavelength is 
\begin{equation}
    \tk_{y,{\rm max}}=\frac{1}{2}+\frac{\pi \tml}{16\tw},
    \label{eq:kmaxshort1}
\end{equation}
and for $\tk_y \gg 1$,
\begin{equation}
    \tk_{y,{\rm max}}= \left(\frac{\pi\tml}{2\tw} \right)^{1/3}.
    \label{eq:kmaxshort2}
\end{equation}
We set the boundary between these two approximate solutions at the point where these two equations have the same values, which are $\tk_y\simeq 2.5$ with $\tml/\tw\simeq 10$.
We can obtain the instability condition from the most unstable wavelength.
Using Equation (\ref{eq:kmaxshort1}), we obtain ${\tilde \omega}^2=0$ for $\tml\simeq 1.3\tw$, which is the instability condition for $\tk_y\tw \gtrsim 2$.
As $\tml/\tw=2/(\pi Q)$, this condition can be rewritten as $Q\simeq 0.49$.
This is slightly different from the condition for the GI of an infinitesimally thin disk, $Q=1$, as  we consider the effect of disk thickness here.

As shown in the bottom panel of Figure \ref{fig:param_surv}, $\tk_{y,{\rm crit}}$ is smaller than $\sim 1$.
Thus, to estimate $\tk_{y,{\rm crit}}$, we assume $\tk_{y,{\rm crit}} \ll 1$ and replace $\tk_y/(1+\tk_y)$ with $\tk$ in the second term on the left-hand side of Equation (\ref{eq:disp_short}).
Consequently, we obtain
\begin{equation}
    \tk_{y,{\rm crit}}=\frac{\pi \tml}{2 \tw}-\sqrt{\left(\frac{\pi \tml}{2\tilde{W}}\right)^2-1}.
\end{equation}
Since ${\tilde k}_{y,{\rm crit}}$ is required to be in the range ${\tilde k} _y\tw\gtrsim 2$, $\tml$ and $\tw$ satisfy the relation 
\begin{equation}
 \tml<\frac{\tw^2+4}{2\pi}.
 \label{eq:kcritshortbound}
\end{equation}

\subsubsection{Intermediate Wavenumber \texorpdfstring{$(0.5 \lesssim {\tilde k}_y \tw \lesssim 2)$}{0.5<kW<2}}

For intermediate wavenumbers, we can use the approximation $xg(x)\sim 4/5 x^{1/3}$.
The approximated dispersion relation then becomes
\begin{equation}
 {\tilde \omega}^2 = {\tilde k}_y^2 -\frac{4\pi }{5} \tml \tw^{-2/3}
\frac{{\tilde k}_y^{4/3}}{1+\tk_y}+1.
\label{eq:disp_mid}
\end{equation}
In this case, we use the approximations $\tk_y\gg 1$ and $\tk_y\ll1$ to evaluate $\tk_{y,{\rm max}}$ and $\tk_{y,{\rm crit}}$, respectively, and we obtain 
\begin{equation}
    \tk_{y,{\rm max}} = \left( \frac{2\pi\tml}{15\tw^{2/3}}\right)^{3/5},
    \label{eq:kmaxmid}
\end{equation}
and 
\begin{equation}
    \tk_{y,{\rm crit}} = \left(\frac{5}{4 \pi \tml}\right)^{3/4}\tw^{1/2}.
    \label{eq:kcritmid}
\end{equation}
To evaluate $\tk_{y,{\rm crit}}$, we neglect the term $\tk_y^2$ in Equation (\ref{eq:disp_mid}) and replace $\tk_y^{4/3}/(1+\tk_y)$ with $\tk_y^{4/3}$. 
Using Equation (\ref{eq:kmaxmid}), we then obtain the instability condition $\tml\simeq1.7\tw^{2/3}$.
The condition where ${\tilde k}_{y,{\rm crit}}$ is in the range of $0.5\lesssim {\tilde k_y}\tw \lesssim 2$ is given by 
\begin{equation}
 0.16\tw^2\lesssim \tml \lesssim \tw^2.
 \label{eq:kcritmidbound}
\end{equation}

\subsubsection{Small Wavenumber \texorpdfstring{$({\tilde k}_y\tw \lesssim 0.5)$}{kW<0.5}}

For the small wavenumber case, we can use the approximation $xg(x)\sim x^{3/4}$ to obtain the  dispersion relation as
\begin{equation}
 {\tilde \omega}^2 = {\tilde k}_y^2 -\pi \tml \tw^{-1/4} \frac{\tk^{7/4}}{1+\tk}+1.
 \label{eq:disp_long}
\end{equation}
Using the approximations $\tk_{y,{\rm max}}\gg 1$ and $\tk_{y,{\rm crit}}\ll1$, we evaluate the approximate formulas and obtain 
\begin{equation}
    \tk_{y,{\rm max}} = \left( \frac{3\pi\tml}{8\tw^{1/4}}\right)^{4/5}
    \label{eq:kmaxlong}
\end{equation}
and 
\begin{equation}
    \tk_{y,{\rm crit}} = \frac{\tw^{1/7}}{(\pi \tml)^{4/7}}.
    \label{eq:kcritlong}
\end{equation}
To evaluate $\tk_{y,{\rm crit}}$, we neglect the term $\tk_y^2$ and replace $\tk_y^{7/4}/(1+\tk_y)$ with $\tk_y^{7/4}$ in Equation (\ref{eq:disp_long}). 
From Equation (\ref{eq:kmaxlong}), the instability condition is $\tml\simeq 1.2\tw^{1/4}$.
If ${\tilde k}_{y,{\rm crit}}$ is in the range of ${\tilde k}_y\tw < 0.5$, the parameters $\tml$ and $\tw$ satisfy the relation
\begin{equation}
 \tml\gtrsim \tw^2.
 \label{eq:kcritlongbound}
\end{equation}

\subsubsection{Boundaries of the Parameter Region}

We define the boundary in parameter space between the equations for $\tk_{y,{\rm max}}$ for large (Equation (\ref{eq:kmaxshort1})) and intermediate (Equation (\ref{eq:kmaxmid})) wavenumbers by the condition that Equation (\ref{eq:kmaxmid}) satisfies $\tk_y\tw=2$.
The boundary for $\tk_{y,{\rm max}}$ between intermediate (Equation (\ref{eq:kmaxmid})) and small (Equation (\ref{eq:kmaxlong})) wavenumbers is given by the condition that Equation (\ref{eq:kmaxlong}) satisfies $\tk_y\tw=0.5$;
these boundaries are $ \tml=7.6\tw^{-1}$ and $\tml=1.1\tw$, respectively.

The boundary for $\tk_{y,{\rm crit}}$ is given by Equations (\ref{eq:kcritshortbound}), (\ref{eq:kcritmidbound}), and (\ref{eq:kcritlongbound}).
Equation (\ref{eq:kcritshortbound}) and the left-hand side of Equation (\ref{eq:kcritmidbound}) are different and the region where the parameters satisfy both Equations (\ref{eq:kcritshortbound}) and (\ref{eq:kcritmidbound}) appears because we use a different way for the approximation of the equations in the case of intermediate and small wavenumber. 
In such regions, we can use either approximation. 
They have similar values for large $\tml$ or $\tw$ and the boundary is roughly given by  $\tw\sim \sqrt{6 \tml}$.
We use this boundary for the comparison with the numerical results.
We plot the boundaries $\tw=\sqrt{6\tml}$ and $\tw = \sqrt{\tml}$ in the bottom panel of Figure \ref{fig:param_surv} (the dotted and dashed lines, respectively).
Because the line $\tw=\sqrt{6\tml}$ is in the unstable region for $\tw\gtrsim 10$ and $\tml \gtrsim 10 $, this approximation is justified.

We test the approximated equations with the numerical results and find that the error is within about 50\% in the unstable region shown in Figure \ref{fig:param_surv} (see Appendix \ref{deri}).

\section{GI of rings formed by the secular GI}
\label{result}

\subsection{Ring Model}
\label{RingModel}

In this section, we investigate the process in which a dust ring structure is formed by the secular GI and the ring collapses due to the ring GI to form a planetesimal or planet.
When the secular GI grows enough and starts the nonlinear growth, it transitions to a GI-like contraction in the radial direction whose timescale is similar to the dynamical timescale. If the conditions for ring GI are satisfied after the transition to dynamical radial GI-like contraction, the ring GI can grow after the saturation of the dynamic radial gravitational collapse. Thus, we can treat this process as a continuous and independent process of the secular GI and the ring GI.

\begin{figure}[tb]
\includegraphics[width=8cm]{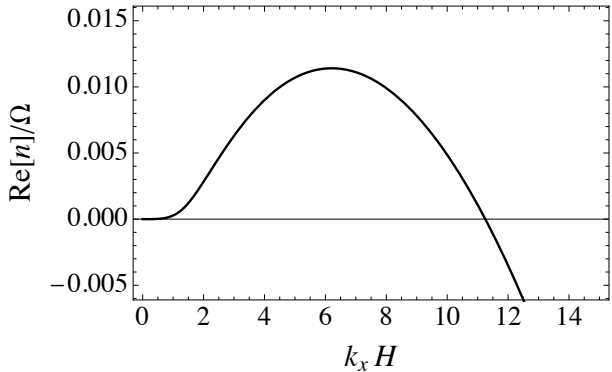}
\caption{
Dispersion relation for a self-gravitating gas-dust disk with $\alpha=3\times10^{-4},\ \epsilon=0.1,\ Q=3$, and St=0.1.
The horizontal axis is the normalized wavenumber where $k_y$ is the perturbation wavenumber  and $H$ is the scale height of the gas. The vertical axis is the normalized growth rate, where $n$ is the growth rate and $\Omega$ is the Kepler frequency.
}
\label{fig:SGI}
\end{figure}

We use the basic equations obtained in \cite{2019ApJ...881...53T} to obtain the dispersion relation for a self-gravitating gas-dust disk.
The secular GI is controlled by four parameters: the turbulent strength $\alpha$ \cite[]{1973A&A....24..337S}, the dust-to-gas mass ratio $\epsilon$, the Toomre parameter $Q$ \cite[]{1964ApJ...139.1217T}, and the Stokes number St.
For simplicity, we assume that these parameters are constant in the disk.
For example, we adopt $\alpha=3\times10^{-4},\ \epsilon=0.1,\ Q=3$, and St=0.1.
We discuss the effects of these disk parameters in Section \ref{sgiparam}.
The dispersion relation for a self-gravitating gas-dust disk is shown in Figure \ref{fig:SGI}.
According to this dispersion relation, the most unstable wavelength reflects the scale height of the gas.

\begin{figure}[tb]
\includegraphics[width=8cm]{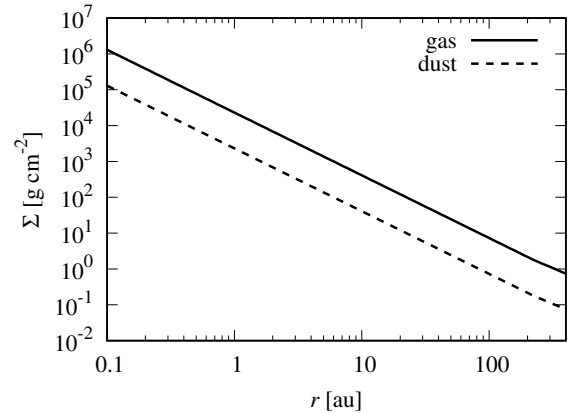}
\caption{Surface density distributions of the gas and dust in a disk model with $\epsilon=0.1,\ Q=3$, $M_*=M_\odot$, and $T(r)={\rm max}[150(r/1 {\rm [au]})^{-1/2},10][{\rm K}]$.}
\label{fig:sigma}
\end{figure}

We consider the ring structure formed by the growth of the perturbation with the most unstable wavelength of the secular GI.
We evaluate the line mass of the ring by multiplying the surface density with the most unstable wavelength.
We evaluate the surface density from the assumptions that the mass of the central star is $M_*=M_\odot$ and the temperature distribution is $T(r)={\rm max}[150(r/1 {\rm [au]})^{-1/2},10][{\rm K}]$.
Using these assumptions, we obtain the surface density of the gas to be $\Sigma = \cs\Omega/(\pi G Q)$.
Figure \ref{fig:sigma} shows the surface density of the gas and dust for this disk model, where we evaluate the surface density of the dust as $\Sigma_{\rm d}=\epsilon \Sigma$.
The line mass of the dust ring formed by the secular GI is given by $M_{\rm L}=\Sigma_{\rm d}\lambda_{\rm SGI}$, where $\lambda_{\rm SGI}$ is the most unstable wavelength of the secular GI.

For the ring GI, we need the velocity dispersion of the dust and the ring width in addition to the line mass of the ring obtained from the stability analysis of the secular GI.
However, the velocity dispersion and the ring width are highly uncertain.
The ring width is determined by the time at which the growth of the secular GI has saturated, while the velocity dispersion depends on the collisions among the dust grains, the growth of dust in the ring structure, and the gravitational energy released in the nonlinear growth of the secular GI.
In this study, we introduce parameters for the ring width and velocity dispersion.
We introduce the parameter $e_{\rm d}$, which is the ratio of the velocity dispersion to the Kepler velocity, 
\begin{equation}
e_{\rm d}=\cd/(r\Omega).
\end{equation}
Using $e_{\rm d}$, we can present the ring width in the form $W=\tw H_{\rm d}=\tw e_{\rm d}r$.
In this section, we set $\tw=1$ and show the results using some $e_{\rm d}$ values. 
We discuss the $\tw$ dependence in Section \ref{wdepend}.

\begin{figure}[tb]
\includegraphics[width=8cm]{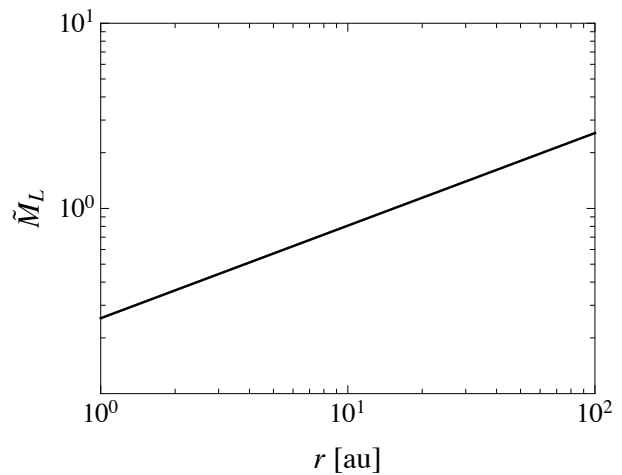}
\caption{Radial distribution of $\tml$ for $e=0.005$.}
\label{fig:f}
\end{figure}

We can evaluate $\tml$ when $e_{\rm d}$ is provided.
The radial distribution of $\tml$ for $e_{\rm d}=0.005$ is shown in Figure \ref{fig:f}.
In this case, $\tml=2$ at $r=61.3$ au and the dispersion relation is given in Figure \ref{fig:ring_disp}.
Because $\tml\propto e_{\rm d}^{-2}$, $\tml$ is large for small $e_{\rm d}$ value, which corresponds to a thin ring.

\subsection{Radial Distribution of the Planetesimal Mass}
\label{calculation_result}

\begin{figure}[tb]
\begin{minipage}{80mm}
\begin{center}
\plotone{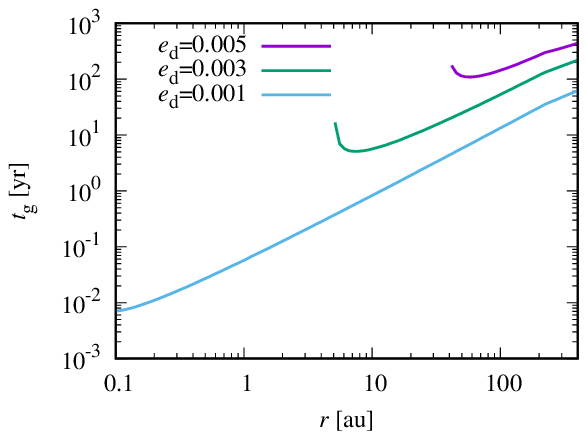}
\end{center}
\end{minipage}
\\
\begin{minipage}{80mm}
\begin{center}
\plotone{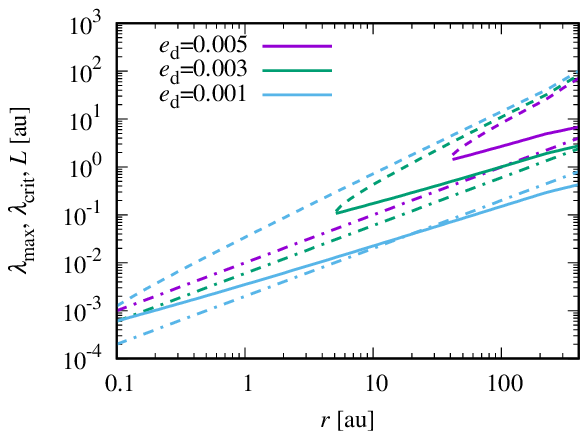}
\end{center}
\end{minipage}
\caption{
Radial distributions of the growth timescale $t_{\rm g}$ at the most unstable wavelength (top panel) and the wavelengths used for mass estimations (bottom panel).
}
\label{fig:thick}
\end{figure}

We performed a linear stability analysis using the corresponding $\tml$ at each radius.
The top panel of Figure \ref{fig:thick} shows the radial distribution of the growth timescale of the ring GI at ${\tilde k}_{\rm y,max}$. The bottom panel of Figure \ref{fig:thick} shows the most unstable wavelength, the largest unstable wavelength, and the ring width.
In both panels, we show the results with $e_{\rm d}=0.005,\ 0.003$, and $0.001$.
The outer part of the disk is more prone to being unstable, and the smaller the $e_{\rm d}$ value, the larger the region that becomes unstable.
For the case with  $e_{\rm d} = 0.01$, the disk is stable in the entire region of 0.1-400 au calculated here.
The growth timescale decreases from the outside to the inside of the disk, and it increases again around the radius of the boundary where the ring GI becomes stable.
For $e_{\rm d} = 0.005$, the growth time is $\sim 400$ yr at 400 au and the instability grows faster than the dynamical time, except in the marginally unstable region. 
The solid lines in the bottom panel show the most unstable wavelengths;
the smaller the radius and $e_{\rm d}$ values , the smaller the most unstable wavelength. 
The dashed lines show the largest unstable wavelengths. 
The slopes are steeper than those of the most unstable wavelengths.
Although the largest unstable wavelength increases as $e_{\rm d}$ decreases, its dependence on $e_{\rm d}$ is smaller than that for the most unstable wavelength.
The dash-dotted lines represent the ring width ($2e_{\rm d}r$).
While the largest unstable wavelength is sufficiently larger than the ring width, the most unstable  wavelength is smaller than the width of $r\gtrsim20$ with $e_{\rm d} = 0.001$. 
We must be careful about the results in this region because the linear stability analysis is based on the thin ring approximation.

\begin{figure}[tb]
\includegraphics[width=80mm]{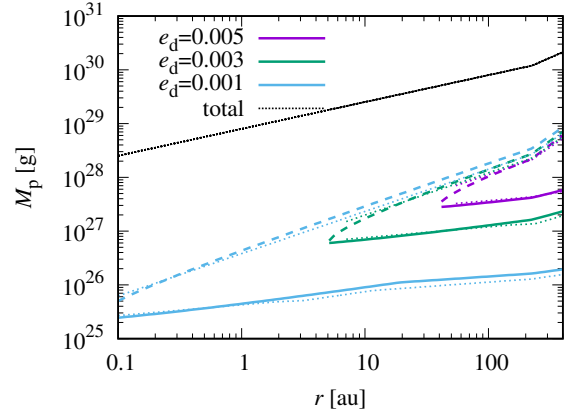}
\caption{
Estimated upper and lower limits of the planetesimal mass. The colored solid lines show the lower limits of obtained from the most unstable wavelength, and the dashed lines show the upper limits obtained from the largest unstable wavelength. 
The dotted lines show the approximate solution obtained in Section \ref{app_mass}.
The black solid line shows the total dust mass of the ring.
}
\label{fig:mass_thick}
\end{figure}

In the case where a ring fragments and forms planetesimals, the ring is expected to fragment first at the most unstable wavelength $\lambda_{\rm max}$, where it shows the fastest growth. 
Then, perturbations with longer wavelengths can grow to form larger planetesimals. 
Therefore, we consider the planetesimal mass $M_{\rm p}$ given by the most unstable wavelength to be the lower limit and the mass $M_{\rm p}$ given by the largest unstable wavelength $\lambda_{\rm crit}$ to be the upper limit of the planetesimal mass.
Figure \ref{fig:mass_thick} shows these upper and lower limits of the planetesimal mass estimated for each value of $e_{\rm d}$.
To estimate the mass, we use $M_{\rm L}\lambda$ when the wavelength is larger than the full width of the ring $L=2W$, and $M_{\rm L}\lambda_{\rm ring}^2/L$ when the wavelength is smaller than $L$.
We also plot the total dust mass of the ring in Figure \ref{fig:mass_thick} (the black solid line). 
This gives the maximum mass formed of the solid body in the single ring.
The upper limit of the mass (dashed lines) shows that highly massive bodies, exceeding $10^{28}$ g, can undergo direct gravitational collapse due to the ring GI beyond 100 au.

\subsection{Approximate Solutions for the Planetesimal Mass}
\label{app_mass}

We analytically evaluate the approximate upper and lower limits for the radial distribution of the planetesimal mass.

\subsubsection{Line Mass of the Ring Formed by the Secular GI}

To obtain $\tml$, we first need to determine the most unstable wavelength of the secular GI.
For simplicity, we neglect the back-reaction from dust to gas and use the approximations $\alpha\ll 1$, $\St\ll 1$, and $n/\Omega\gg 1$.
In this case, the dispersion relation for a self-gravitating gas-dust disk is 
\begin{equation}
 n\simeq-Dk_x^2+2 \pi G \epsilon \Sigma t_{\rm stop}
\frac{k_x}{1+k_x\sqrt{\alpha/\St}H_{\rm g}},
\label{eq:SGIapp}
\end{equation}
where $k_{\rm x}$ is the wavenumber in radial direction and $D=\alpha\cs^2/\Omega$ is the diffusion coefficient of the dust due to turbulence. 
The basic equations used to derive the Equation (\ref{eq:SGIapp}) are given by Appendix \ref{appen}.
The most unstable wavelength can be obtained from $\partial n/\partial k_x=0$, and it satisfies the relation  
\begin{equation}
 k_xH_{\rm g}(1+k_x\sqrt{\alpha/St}H_{\rm g})^2=\frac{\epsilon {\rm St}}{Q \alpha}
\end{equation}
For the long-wavelength limit ($k_x\sqrt{\alpha/\St}H_{\rm g}\ll 1 $), the most unstable wavenumber is 
\begin{equation}
 k_{\rm SGI,l} \simeq \frac{\epsilon \St}{Q \alpha}H_{\rm g}^{-1}.
\label{eq:kSGIl}
\end{equation}
For the short wavelength limit ($k_x\sqrt{\alpha/\St}H_{\rm g}\gg 1 $), the most unstable wavenumber is 
\begin{equation}
 k_{\rm SGI,s} \simeq \left(\frac{\epsilon}{Q}\right)^{1/3}\left(\frac{\St}{\alpha}\right)^{2/3}H_{\rm g}^{-1}.
\label{eq:kSGIs}
\end{equation}
Thus, we evaluate the most unstable wavenumber approximately as 
\begin{equation}
 k_{\rm SGI} = 
\left( k_{\rm SGI,l}^{-1} + k_{\rm SGI,s}^{-1} \right)
^{-1}.
\label{eq:kSGI}
\end{equation}

Using this most unstable wavenumber, we can obtain $\tml$ as
\begin{equation}
 \tml=\frac{G \Ml}{\cd^2}=\frac{G \Sigma_{\rm d}}{\cd^2}\frac{2\pi}{k_{\rm SGI}}
=\frac{2\epsilon}{Qe_{\rm d}^2}(rk_{\rm SGI})^{-1}\left(\frac{H_{\rm g}}{r}\right).
\label{eq:tml_sgi}
\end{equation}

\subsubsection{Lower and Upper limits of the Planetesimal Mass}
\label{appmax}

From Equation (\ref{eq:tml_sgi}) and the approximated formulas for $\tk_{y,{\rm max}}$ and $\tk_{y,{\rm crit}}$ obtained in Section \ref{app}, we can evaluate the lower and upper limits of the planetesimal mass.
Below, We give examples of the approximate mass distribution formed by the ring GI.

In the case where $\lambda_{\rm max}>L$, $k_{\rm SGI}=k_{\rm SGI,l}$, and $\tk_{y,{\rm max}}$ given by Equation (\ref{eq:kmaxshort2}), we obtain the lower limit of $M_{\rm p}$ as
\begin{eqnarray}
 M_{\rm p} &=& M_{\rm L}\lambda_{\rm max} \nonumber \\
&\simeq&
8.6\left( \frac{\alpha}{\St} \right)^{2/3}
e_{\rm d}^{5/3}w^{1/3}
\left(\frac{H}{r}\right)^{4/3}
M_{\rm s} \nonumber\\
&\simeq&
0.4M_{\rm \oplus}
\left(\frac{\alpha}{3\times10^{-4}}\right)^{2/3}
\left(\frac{{\rm St}}{0.1}\right)^{-2/3}\nonumber\\
&&\left(\frac{e_{\rm d}}{0.005}\right)^{5/3}
\left(\frac{H_{\rm g}/r}{0.1}\right)^{4/3}
w^{1/3}
\left(\frac{M_{\rm s}}{M_{\rm \odot}}\right)
\label{eq:appMmin}
\end{eqnarray}

For the upper limit of the planetesimal mass, we use Equation (\ref{eq:kcritlong}) to obtain the largest unstable wavelength.
For $\lambda_{\rm crit}>L$ and $k_{\rm SGI}=k_{\rm SGI,l}$, the mass $M_{\rm p}$ is given by
\begin{eqnarray}
 M_{\rm p} &\simeq& 36 
\left( \frac{\alpha}{\St} \right)^{11/7}
e_{\rm d}^{-1/7}w^{-1/7}
\left(\frac{H_{\rm g}}{r}\right)^{22/7}
M_{\rm s} \nonumber\\
&\simeq&
2.0M_{\rm \oplus}
\left(\frac{\alpha}{3\times10^{-4}}\right)^{11/7}
\left(\frac{{\rm St}}{0.1}\right)^{-11/7}\nonumber\\
&&\left(\frac{e_{\rm d}}{0.005}\right)^{-1/7}
\left(\frac{H_{\rm g}/r}{0.1}\right)^{22/7}
w^{-1/7}
\left(\frac{M_{\rm s}}{M_{\rm \odot}}\right).
\label{eq:appMmax}
\end{eqnarray}

These lower and upper limits of the planetesimal mass are presented by the dotted curves in Figure \ref{fig:mass_thick}.
The approximate solutions well reproduce the numerical solutions except at the boundary of the unstable region, where the upper limit is overestimated by a factor of approximately two.
However, within a sufficiently unstable region, we can estimate the planetesimal mass using the approximate solution.

\subsection{Parameter Dependence of the Planetesimal Mass}

In this section, we discuss the dependence of the planetesimal mass on the parameter $\alpha,\ \St,\ Q,\ \epsilon$ and $\tw$.
We have investigated the masses of the planetesimals formed by the ring GI for some $e$ values, fixed disk parameters $(\alpha,\ \St,\ Q,\ \epsilon)=(3\times10^{-4},\ 0.1,\ 3,\ 0.1)$, and the ring width parameter $\tw=1$ in Section \ref{calculation_result}.
These masses change when we use the different parameters as expected from the approximate formula (Equations (\ref{eq:appMmin}) and (\ref{eq:appMmax})).

\subsubsection{Dependence of the Planetesimal Mass on the Disk Parameters}
\label{sgiparam}
First, we show the radial distributions of the growth timescales, the most unstable and the largest unstable wavelengths and the planetesimal masses for different $\alpha$ and $\St$.
Here, we adopt $(Q,\ \epsilon)=(3,\ 0.1)$, $e_{\rm d}=0.001$, and  $\tw=1$,
and we compare the results with $(\alpha,\ \St)=(3\times10^{-4},\ 0.1),\ (10^{-5},\ 0.1)$, and $(10^{-5},\ 0.01)$.
The approximate equations (\ref{eq:kSGIl}), (\ref{eq:kSGIs}), and (\ref{eq:kSGI}) show that the most unstable wavelength of the secular GI depends on $\alpha/\St$ and it becomes shorter for the smaller $\alpha$ and larger St values.
Therefore, the model with $(\alpha,\ \St)=(10^{-5},\ 0.1)$  has a smaller values of the most unstable wavelength than that with $(\alpha,\ \St)=(3\times10^{-4},\ 0.1)$, and the model with $(\alpha,\ \St)=(10^{-5},\ 0.01)$ has the most unstable wavelength between these two. 
The most unstable wavelength of the secular GI affects the ring GI through the line mass of the ring.
The shorter the wavelength, the smaller the line mass of each ring and $\tml$.

\begin{figure}[tb]
\begin{minipage}{80mm}
\begin{center}
\plotone{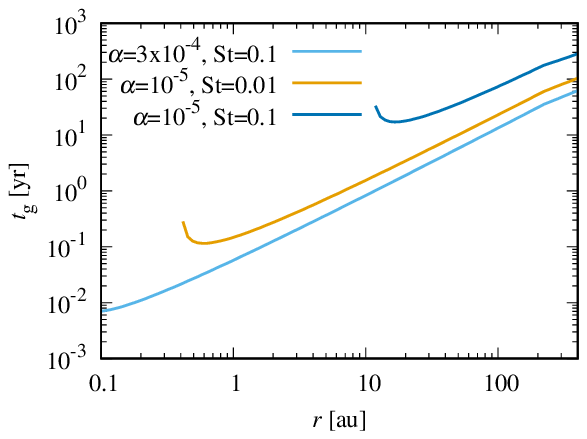}
\end{center}
\end{minipage}
\\
\begin{minipage}{80mm}
\begin{center}
\plotone{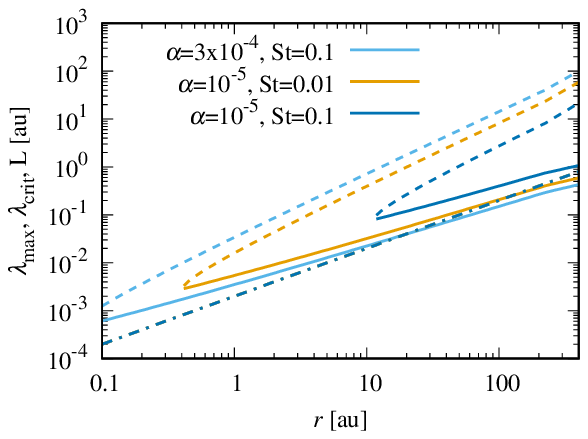}
\end{center}
\end{minipage}
\\
\begin{minipage}{80mm}
\begin{center}
\plotone{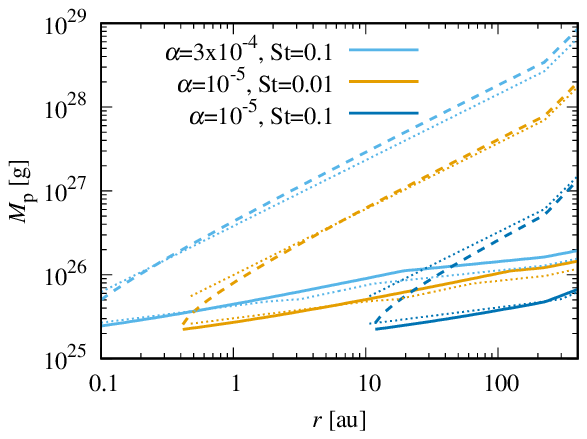}
\end{center}
\end{minipage}
\caption{
Comparison of the growth timescales, key wavelengths, and planetesimal masses for $(\alpha,\ \St)=(3\times10^{-4},\ 0.1),\ (10^{-5},\ 0.1)$, and $(10^{-5},\ 0.01)$.
The other parameters are $(Q,\ \epsilon)=(3,\ 0.1)$, $e_{\rm d}=0.001$, and $\tw=1$.
}
\label{fig:comp_alpha_St}
\end{figure}

The upper panel of Figure \ref{fig:comp_alpha_St} shows the comparison of the growth timescales of the three models at the most unstable wavelength.
The smaller the line mass of the ring, the longer the growth timescale and the larger the radius of the inner boundary of the unstable region. 
The middle panel of Figure \ref{fig:comp_alpha_St} shows the comparison of the wavelengths $\lambda_{\rm max}$ and $\lambda_{\rm crit}$ for three models.
The ring width is the same in all three models because the ring's scale heights and $\tw$ are the same.
This figure shows that the shorter the most unstable wavelength of the secular GI, the longer the most unstable wavelength and the shorter the largest unstable wavelength of the ring GI.
 These trends are similar to those observed in Section \ref{calculation_result}, when $e_{\rm d}$ is varied.
This is because the non-dimensional dispersion relation given by Equation (\ref{eq:disp_norm}) is affected by the change in $\tml$  when the line mass of the ring or $e_{\rm d}$ is changed.
The bottom panel of Figure \ref{fig:comp_alpha_St} shows the comparison of the masses of the planetesimals formed by the ring GI.
It shows that both the upper and lower limits of the mass decrease as the line mass of the ring decreases.
The upper limit is more dependent on the line mass of the ring than the lower limit.
These planetesimal masses are given by multiplying the line mass with the wavelength.
The upper mass limit decreases because both the line mass and the largest unstable mass decrease.
Meanwhile, as the most unstable wavelength increases as the line mass of the ring decreases, the lower mass limits are less dependent on the line mass than the upper limits.

Next, we discuss the cases in which $Q$ and $\epsilon$ are varied. 
When the wavelength is sufficiently long,
$k_{\rm SGI,l}\propto \epsilon/Q\propto\Sigma_{\rm d}$ (Equation (\ref{eq:kSGIl})) and the ring line mass $M_{\rm L}\propto\Sigma_{\rm d}/k_{\rm SGI,l}$ become independent of $\epsilon$ and $Q$. 
Consequently, the approximate equations (\ref{eq:appMmin}) and (\ref{eq:appMmax}) for the upper and lower limits of the planetesimal  masses do not depend on these parameters. 
However, when the most unstable wavelength of the secular GI is modified mainly owing to the effect of  disk thickness, a dependence on $\epsilon$ and $Q$ appears. 
If the wavelength is short enough, $k_{\rm SGI} \sim k_{\rm SGI,s}$  is proportional to $(\epsilon/Q)^{1/3}$ (Equation (\ref{eq:kSGIs})).
In this case, the line mass of the ring increases with increasing $\epsilon/Q$, and a dependence on $\epsilon$ and $Q$ appears. 
Here, we compare the results for $(Q,\ \epsilon)=(3,\ 0.1),\ (10,\ 0.1)$, and $(3,\ 0.01)$, assuming that  $(\alpha,\ \St)=(10^{-5},\ 0.1)$, $e_{\rm d}=0.001$, and  $\tw=1$.

\begin{figure}[tb]
\begin{minipage}{80mm}
\begin{center}
\plotone{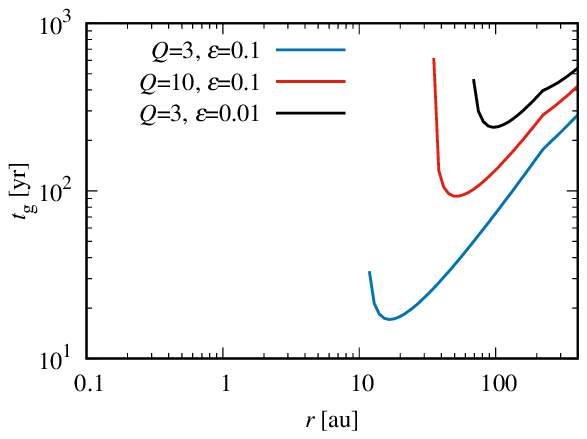}
\end{center}
\end{minipage}
\\
\begin{minipage}{80mm}
\begin{center}
\plotone{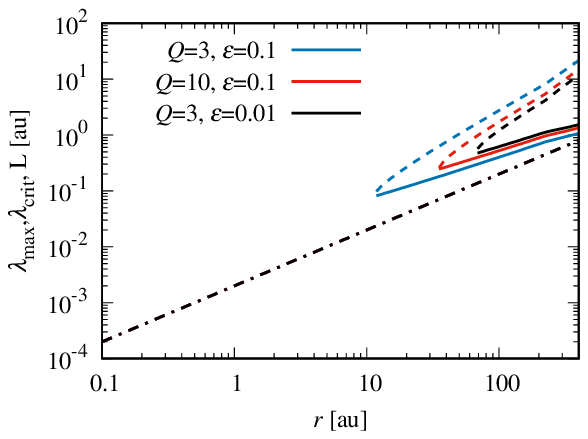}
\end{center}
\end{minipage}
\\
\begin{minipage}{80mm}
\begin{center}
\plotone{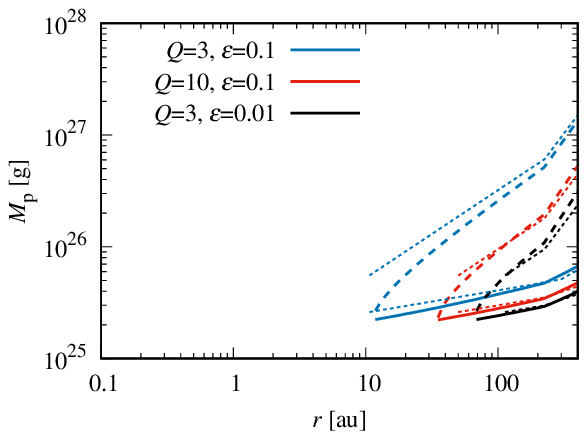}
\end{center}
\end{minipage}
\caption{
Comparison of the growth timescales, key wavelengths, and planetesimal masses for $(Q,\ \epsilon)=(3,\ 0.1),\ (10,\ 0.1)$, and  $(3,\ 0.01)$ with $(\alpha,\ \St)=(10^{-5},\ 0.1)$, $e_{\rm d}=0.001$, and $\tw=1$.
}
\label{fig:comp_Q_epsilon}
\end{figure}

Figure \ref{fig:comp_Q_epsilon} shows the comparison of the growth rate at ${\tilde k}_{y,{\rm max}}$ for these models.
As mentioned above, the parameters of the secular GI affect the ring GI through the ring line mass.
In these models, the line mass of the ring decreases in the order of $(Q,\ \epsilon)=(3,\ 0.1),\ (10,\ 0.1)$, and $(3,\ 0.01)$.
Therefore, the differences between these models are basically the same as the differences that occur when the line mass of the ring decreases when $\alpha$ and St are changed.

In the bottom panels of Figures \ref{fig:comp_alpha_St} and \ref{fig:comp_Q_epsilon}, we show the approximate upper and lower limits of the planetesimal mass (dottede lines).
These panels show that the most unstable wavelength of the secular GI  is well estimated by Equation (\ref{eq:kSGI}), because except the boundary of the unstable region, the numerical solution is well reproduced for all parameters.

\subsubsection{Dependence of the Planetesimal Mass on the Ring Width}
\label{wdepend}

\begin{figure}[tb]
\begin{minipage}{80mm}
\begin{center}
\plotone{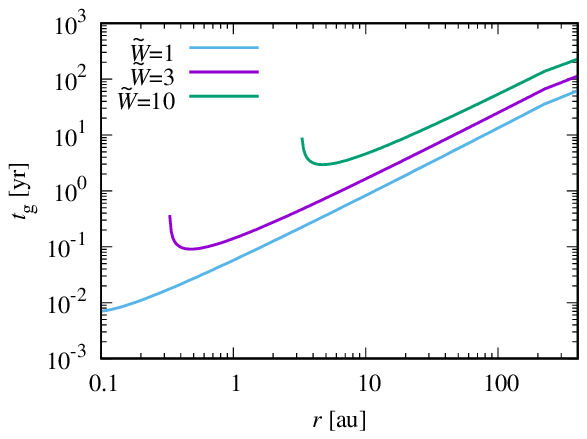}
\end{center}
\end{minipage}
\\
\begin{minipage}{80mm}
\begin{center}
\plotone{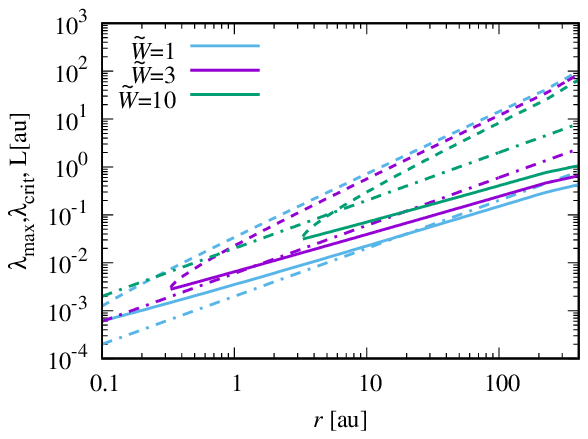}
\end{center}
\end{minipage}
\\
\begin{minipage}{80mm}
\begin{center}
\plotone{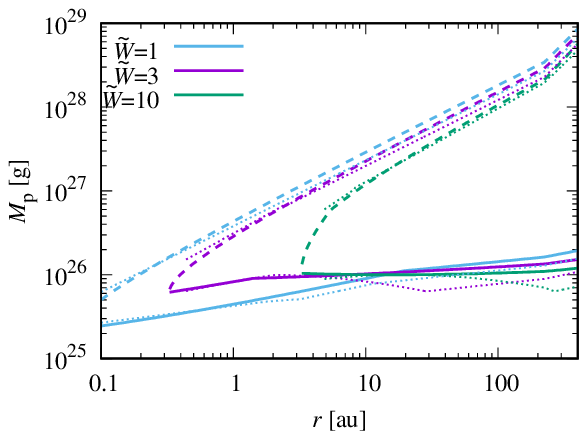}
\end{center}
\end{minipage}
\caption{
Comparison of the growth timescales, key wavelengths, and planetesimal masses for $\tw = 1,\  3$, and  $10$ with $(\alpha,\ \St,\ Q,\ \epsilon)=(3\times10^{-4},\ 0.1,\ 3,\ 0.1)$ and $e_{\rm d}=0.001$.
\label{fig:comp_w}}
\end{figure}

In this section, we consider the case in which the ratio of the width to the thickness of the ring is larger than in models investigated above.
We compare the results for $\tw=1,3$, and $10$, where $(\alpha,\ \St,\ Q,\ \epsilon)=(3\times10^{-4},\ 0.1,\ 3,\ 0.1)$, and $e_{\rm d}=0.001$.
Figure \ref{fig:comp_w} shows the comparison of the three models.
As $\tw$ increases, the surface density of the dust decreases and the gravity becomes weaker. 
Thus, change that occur when $\tw$ increases show tendencies similar to those considered in the previous section, in which the line mass of the ring is decreased.
In contrast to the previous results, the relationship between the unstable wavelength and the ring width changes drastically when $\tw$ is changed. 
At $\tw=1$, the most unstable wavelength exceeds the ring width only above several tens of au, whereas at $\tw = 3$, the most unstable wavelength exceeds the ring width in almost all regions.
In addition, at $\tw = 10$, the largest unstable wavelength is smaller than the ring width near the boundary of the unstable region.
Therefore, the $\tw$ dependence of the planetesimal mass changes in a complicated manner.
According to the approximate equations (\ref{eq:appMmin}) and (\ref{eq:appMmax}), the planetesimal mass does not depend strongly on $\tw$.
The planetesimal mass is approximately proportional to $\tw^{1/3}$ in Equation (\ref{eq:appMmin}) and  to $\tw^{-1/7}$ in Equation (\ref{eq:appMmax}).
The lower limit of the planetesimal mass increases slightly with increasing $\tw$ in the inner region $r\lesssim 10$ au.
However, the dependence is reversed for $r\gtrsim 10$ au because the most unstable wavelength is smaller than the ring width in this region.
This figure shows that the approximate estimates of the planetesimal masses reproduce the results well even when $\tw$ is varied.

\subsection{Expected Direction of Planetesimal Rotation}
\label{spin}

\begin{figure}[tb]
\vspace{5mm}
\includegraphics[width=8cm]{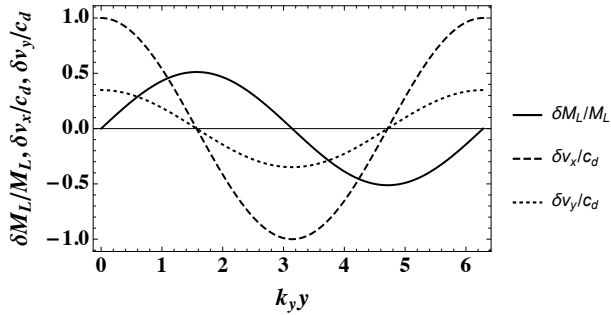}
\caption{Eigenfunctions of $\delta M_{\rm L}$, $\delta v_x$,and  $\delta v_y$ of the ring GI with $\tml=2$, and $\tw=1$, normalized using the maximum value of $\delta v_x/\cd$.
}
\label{fig:eigenvector}
\end{figure}

We discuss the direction of the rotation of the planetesimals formed by the ring GI using the eigenfunctions of the GI.
Figure \ref{fig:eigenvector} shows the eigenfunctions for the most unstable mode of the ring GI with $\tml = 2$ and $\tw = 1$. 
This figure shows that the phases of $\delta v_x$ and $\delta v_y$ are the same and there is a perturbation of the velocity field with positive angular momentum around the gravitationally contracting point where $\delta M_{\rm L}$ is maximum.
Therefore, we expect the planetesimals formed by gravitational collapse caused by the ring GI to have a positive angular momentum.
The origin of this rotation direction is explained by the Coriolis force.
The direction of the velocity field is consistent with the direction of the Coriolis force acting on the dust contracting in the $y$ direction.
Thus, we expect the planetesimals formed by the ring GI to exhibit prograde rotation.

\section{Discussion}
\label{discussion}

\subsection{Fragmentation of the  Rings Formed by the Secular GI} 

However, it could be possible that the ring GI occurs in the linear growth stage of the secular GI.

\cite{2021MNRAS.tmp..225P} performed two-dimensional numerical simulations of the secular GI and  calculated the non-axisymmetric evolution of the ring structures it formed.
In simulations performed by \cite{2021MNRAS.tmp..225P}, the rings formed by the secular GI contract in the azimuthal direction, and the mechanism is considered to be Rossby wave instability, which is based on the Rossby wave instability condition reported by \cite{2017ApJ...849..129L}.
While deriving the instability condition, they assumed the dust to be strongly coupled with the gas.
However, in the present study, we assume that the dust grows sufficiently during ring formation to decouple from the gas.
Therefore, whether the ring fragments by the ring GI or by Rossby wave instability depends on the growth of the dust during the ring formation.

\subsection{Long-term Evolution of Fragment Rotation}

As discussed in Section \ref{spin}, the fragment formed by the ring GI has prograde rotation.
When the fragment collapses, the single planetesimal or planetesimal binary are expected to be formed.
The rotation direction of the fragment affects the rotation direction of the planetesimals and the obliquity of the planetesimal binaries.

When the single planetesimal is formed, it has prograde rotation.
After the planetesimal formation, they can collide with each other over a long timescale. Those collision can provide additional large spin angular momentum onto the planetesimal depending on the collision angle. It is expected that rotation direction is randomized if the planetesimal experiences the collision. 
Thus, our result gives the initial rotation direction of the planetesimals formed by the ring GI.

\begin{figure}[tb]
\includegraphics[width=8cm]{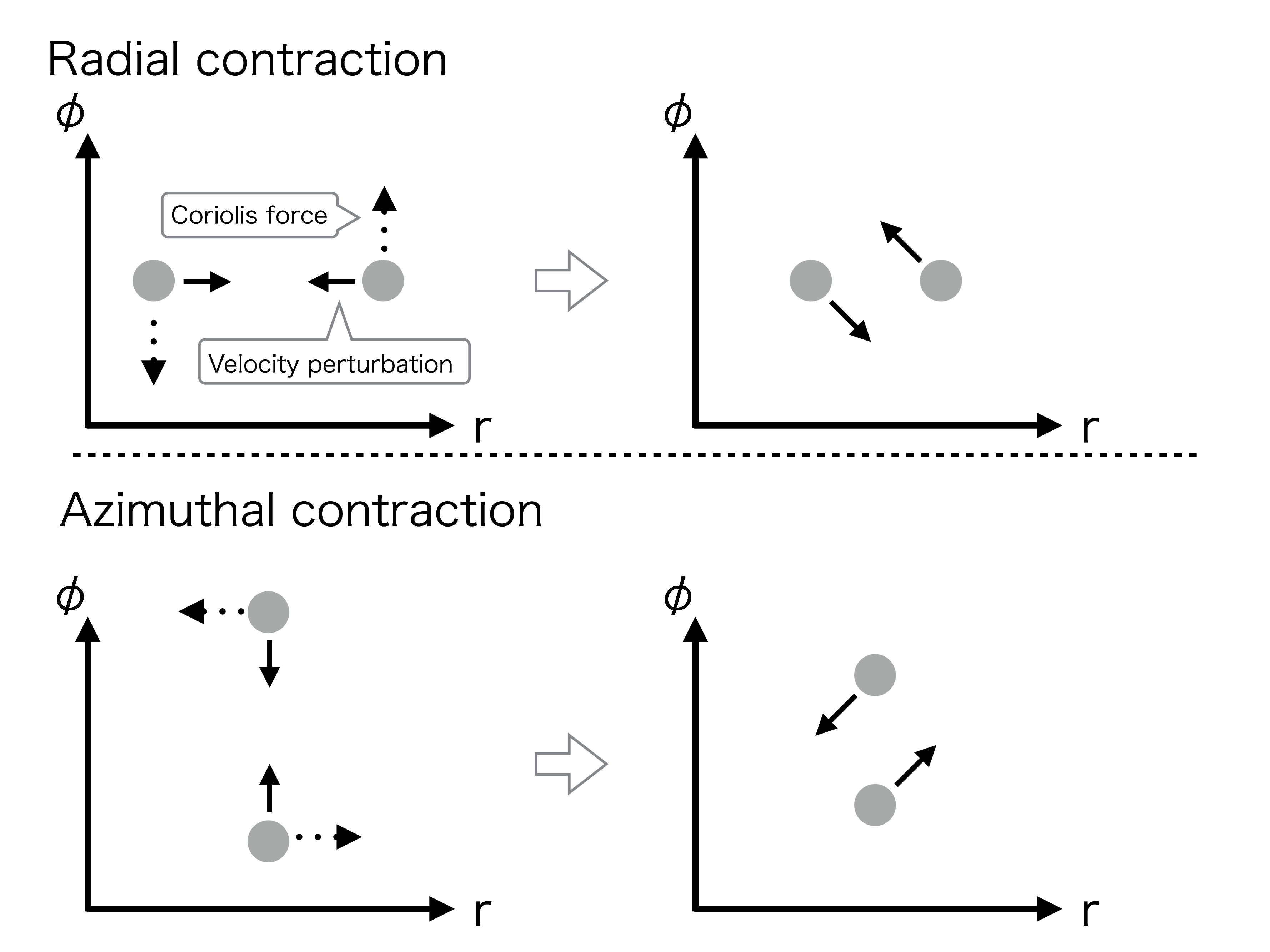}
\caption{Schematic picture showing the rotation direction of the dust contracting in the radial and azimuthal directions. 
The gray filled circles show the dust particles and $r$ and $\phi$ axes show the radial and azimuthal directions, respectively.
}
\label{fig:rotation}
\end{figure}

When the gravitational collapse of the fragment form the planetesimal binary, they should be prograde with zero obliquity.
This nature may explain the obliquity distribution of the trans-Neptunian binaries since the majority of the trans-Neptunian binaries have prograde rotation.
\cite{2019NatAs...3..808N} claim that the obliquity distribution of the Trans-Neptunian binaries is explained by the streaming instability.
They showed that many dust clumps formed by the streaming instability obtain prograde rotation.
They argued that the obliquity distribution of the clumps is consistent with the obliquity distribution of the trans-Neptunian binaries.
When the fragment formed by the GI contracts by the gravitational force, the Coriolis force acts on it, which determines its rotation direction.
For the both radial (GI of the disk) and azimuthal (GI of the ring) contractions, the fragments are expected to obtain prograde rotation by the Coriolis force acting on them (Figure \ref{fig:rotation}).
Thus, the prograde rotation of the trans-Neptunian binaries does not necessarily mean that they are formed by the streaming instability because the prograde rotation can be explained by the gravitational collapse of the fragments without the streaming instability.
It is difficult to directly compare the mass ranges obtained in the previous section and that of Trans-Neptunian binaries, because the lower limit of the mass depends on $e_{\rm d}$ and this parameter is highly uncertain. If the ring is created by a mechanism other than secular GI, the ring line-mass may vary and the mass range of the planetesimal may vary too. Note that our analysis of the ring fragmentation can still be applied to those cases.

Planetesimal binaries can be formed by the gravitational interaction among planetesimals formed in the same ring.
If the planetesimal binaries are created in such a complex dynamics, the resultant rotational directions are not necessarily prograde. 
The planetesimal binaries may experience further interactions with planetesimals over a long timescale. 
It is expected that obliquity direction is randomized by three-body encounters. 
Indeed, some of the observed trans-Neptunian binaries are retrograde, which might be explained by those interactions. 
In other words, the rotation direction distribution and binary obliquity distribution may have the imprint of the dynamical history of fragments after the formation. 
Therefore, further studies of the evolution of fragments should be interesting and will be the subject of our future work.

\subsection{Critical Mass for Ring GI and Its Comparison with the Observations}
The critical mass of the ring for GI is $M_{\rm crit}=2\pi r M_{\rm L,crit}=2 \pi r \tilde{M}_{\rm L,crit}\cd^2/G=2 \pi\tilde{M}_{\rm L,crit} e_{\rm d}^2M_{\rm s}$.
This mass does not depend on the radius but on the central star mass $M_{\rm s}$, normalized critical line mass $M_{\rm L,crit}$ and $e_{\rm d}$.
For $\tilde{W}=1$, we obtain $\tilde{M}_{\rm L,crit}\simeq 1.6$.
The critical mass can be written as
\begin{equation}
    M_{\rm crit}=3.3\times 10^2 M_{\rm \oplus} \left( \frac{\tilde{M}_{\rm L,crit}}{1.6}\right)
    \left( \frac{e_{\rm d}}{0.01}\right)^2
    \left( \frac{M_{\rm \odot}}{M_{\rm s}}\right).
\end{equation}
In the case where the ring width is 1\% of the ring radius, the critical mass is larger than 100$M_{\rm \oplus}$.

Recent observations found the dust ring structures in many protoplanetary disks, with masses in the order of tens of Earth masses and widths of  $\sim$ 10\% of the ring radius \cite[]{2018ApJ...869L..46D}.
Thus, if the dust becomes concentrated and the ring width becomes smaller than 1/10 of the observed width, it may undergo fragmentation through the ring GI.

Notably, we have investigated the gravitational collapse of the dust ring structure without considering its interaction with the gas component of the disk.
This simplification is justified when the dust particle sizes are large and the Stokes number is larger than unity or when the dust-to-gas mass ratio is much larger than unity.
Resent observations at sub-mm wavelengths have found the ring-gap structures in the dust continuum emissions, implying that the dust size in these cases is $\lesssim 1$ mm.
Because such dust particles do not have very large Stokes numbers, the approximation we used to obtain the dispersion relation is not justified in this case.
We plan to investigate the stability of the observed rings, which requires considering the secular GI for such ring structures, for our future study.

\section{Summary}
\label{summary}

We investigated the GI of a dust ring and the resulting planetesimal formation. 
We performed a linear stability analysis of the ring GI, considering the thickness of the ring, and investigated the dependence of the growth timescale, the most unstable wavenumber, and the smallest unstable wavenumber on the ring width and line mass.
We derived approximated formulas that presented the most unstable wavenumber, the smallest unstable wavenumber, and the instability condition.
The error in the approximated wavenumber was less than $\sim$ 50\%.
We found that for the large ring widths and line masses, the instability condition was similar to that of a self-gravitating disk, when the surface density of the ring was given by its line mass divided by the ring width because the most unstable wavelength was smaller than the ring width.
For small ring widths and line masses, however, the most unstable wavelength was larger than the ring width.
In this case, the self-gravity perturbation was smaller than that of the self-gravitating disk, and  the line mass required for the ring GI was larger than that required for a large ring width and line mass.
Moreover, we investigated the planetesimal masses formed by the ring GI when the ring structure was formed by the growth of the secular GI.
We adopted a massive, dust rich disk as the disk model.
We calculated the most unstable wavelength of the secular GI for each disk radius and obtained the line mass of the ring formed through the secular GI by the surface density of the disk and the most unstable wavelength.
We evaluated the lower and upper limits of the planetesimal mass by multiplying the mass of the ring  with the most unstable wavelength and the largest unstable wavelength of the ring GI, respectively.
We found that the upper limit of the planetesimal mass can be as large as $10^{28}$ g. 
Additionally, we found that the ring's width must be smaller than $\sim$1\% of its radius for the ring structure to be gravitationally unstable within $r\lesssim 100$ au.
From the eigenfunctions of the ring GI, we found that there was a perturbation of the velocity field with a positive angular momentum around the gravitationally contracting point.
This indicates that the direction of rotation of the planetesimals formed by the ring GI is likely to be prograde.
This result may be related to the fact that many trans-Neptunian binaries are prograde.
Because the direction of rotation depends on the initial angular momentum and the interactions among the clumps formed by fragmentation, we need the nonlinear simulations and detailed comparison of their results and the observations to clarify the direction of rotation of trans-Neptunian binaries.

\section*{acknowledgements}
This work was supported by JSPS KAKENHI Grant Numbers 18H05436, 18H05437, 18H05438, and 19K14764.
The authors would like to thank Enago (www.enago.jp) for the English language review.

\appendix

\section{Derivation of the approximate formulas for the most unstable wavenumber}
\label{deri}

We derive the approximate formulas for the quantity $\tk_{y,{\rm max}}$, given in Section \ref{app}.
For large wavenumbers $\tk\tw\gtrsim 2$, the dispersion relation is given by Equation (\ref{eq:disp_short}).
The most unstable wavenumber is obtained from $\partial \omega^2/\partial k=0$.
Thus, $\tk_{y,{\rm max}}$ satisfies
\begin{equation}
    \tk_{y,{\rm max}}(1+\tk_{y,{\rm max}})^2=\frac{\pi \tml}{2\tw}.
\end{equation}
In the case where $\tk_{y,{\rm max}}\sim 1$, we set $\tk=1+x$, where $|x|\ll 1$.
Neglecting the higher orders of $x$ we obtain $x=\pi \tml/(16\tw) -1/2$, which can be rewritten as
\begin{equation}
    \tk_{y,{\rm max}} = \frac{1}{2}+\frac{\pi \tml}{16\tw}.
\end{equation}
We obtain the instability condition by substituting this into Equation (\ref{eq:disp_short}) and solving ${\tilde \omega}^2=0$.
Using the relation $\pi \tml/(2\tw)=8x+4$ and neglecting the higher orders of $x$, we obtain the following equation for the normalized frequency at the most unstable wavenumber ${\tilde \omega}_{\rm max}$:
\begin{equation}
    {\tilde \omega}_{\rm max}^2=2-\frac{\pi \tml}{2\tw}.
\end{equation}
From this equation, we find the instability condition to be $\tml/\tw=4/\pi$.

For the intermediate wavenumbers $0.5\lesssim \tk\tw\lesssim 2$, the dispersion relation is given by Equation (\ref{eq:disp_mid}). 
To obtain the most unstable wavenumber, we assume $\tk\ll 1$ and rewrite the dispersion relation as 
\begin{equation}
    {\tilde \omega}^2 = {\tilde k}^2 -\frac{4\pi }{5} \tml \tw^{-2/3}{\tilde k}^{4/3}+1.
\end{equation}
From $\partial \omega^2/\partial k=0$, we obtain
\begin{equation}
    \tk_{y,{\rm max}} = \left( \frac{2\pi\tml}{15\tw^{2/3}}\right)^{3/5}.
    \label{eq:kmaxmid_ap}
\end{equation}
Around the marginally unstable region of parameter space, we have $\tk_{y,{\rm max}}\sim 1$ and the assumption $\tk\ll 1$ is not justified. 
However, we find that the approximate equation obtained under the assumption $\tk\ll 1$ can also mimic $\tk_{y,{\rm max}}$ around the marginally unstable region given by the numerical calculations.
Thus, we use Equation (\ref{eq:kmaxmid_ap}) for the entire parameter space of $0.5\lesssim \tk\tw\lesssim 2$.
The instability condition is derived by solving ${\tilde \omega}^2=0$ with $\tk=\tk_{y,{\rm max}}$.
We introduce $\tk_{y,{\rm max}}=1+x$ and use the fact $|x|\ll 1$ around the boundary.
Neglecting the higher orders of $x$, we obtain the instability condition $\tml\simeq 1.7\tw^{2/3}$.

For the small wavenumbers $\tk\tw \lesssim 0.5$, the dispersion relation is given by Equation (\ref{eq:disp_long}).
We obtain the most unstable wavenumber and the instability condition in a manner similar to that for the case of intermediate wavenumbers.
With the assumption of $\tk\ll 1$, we obtain
\begin{equation}
    \tk_{y,{\rm max}} = \left( \frac{3\pi\tml}{8\tw^{1/4}}\right)^{4/5}.
\end{equation}
Using this approximate function and the fact that $\tk_{y,{\rm max}}\sim 1$ around the marginally unstable region, we obtain the instability condition: $\tml\simeq 1.2\tw^{1/4}$.

Because we use different approximations for these three wavenumber ranges, the parameter boundaries for $\tk_{y,{\rm max}}\tw=2$ obtained for the large and intermediate wavenumbers are different.
The boundaries for $\tk_{y,{\rm max}}\tw=0.5$ obtained for the intermediate and small wavenumbers are also different.
For simplicity, in this study, we adopt the boundary $\tk_{y,{\rm max}}\tw=2$ for the intermediate wavenumbers and $\tk_{y,{\rm max}}\tw=0.5$ for the small wavenumbers.

\begin{figure}[tb]
\includegraphics[width=8cm]{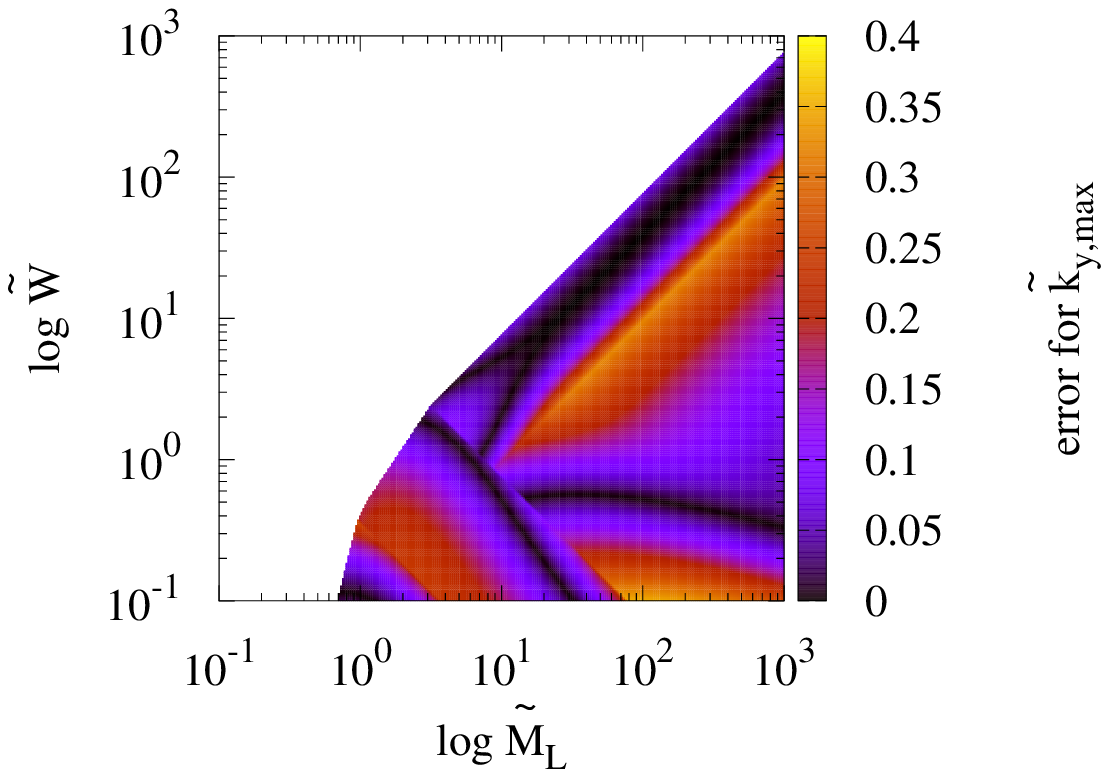}
\includegraphics[width=8cm]{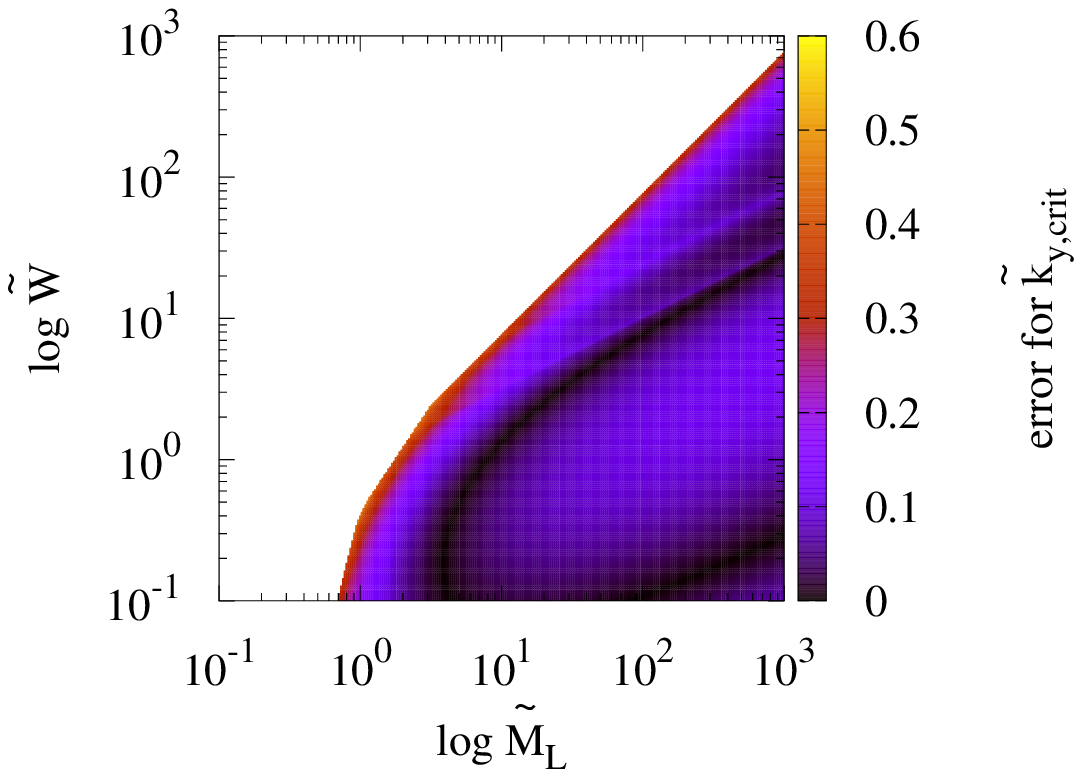}
\caption{Errors in the approximate quantities $\tk_{y,{\rm max}}$ and $\tk_{y,{\rm crit}}$ obtained from the comparison of the results of the approximate formulas with the numerical results. The errors are less than 40\% for $\tk_{y,{\rm max}}$ and 50\% for $\tk_{y,{\rm crit}}$.}
\label{fig:error}
\end{figure}

We compare the results obtained from the numerical calculations with the approximate formulas.
We calculate the errors in the approximate values $|\tk_{\rm app}-\tk_{\rm num}|/\tk_{\rm num}$ when the ring is unstable for both the numerical results and the approximate model with $0.1 \leq \tml \leq 10^3 $ and $0.1 \leq \tw \leq 10^3 $, where $\tk_{\rm app}$ and $\tk_{\rm num}$ are the wavenumbers obtained from the approximate formula and the numerical calculation, respectively.
We show the results in Figure \ref{fig:error}.
We observe that the error for $\tk_{y,{\rm max}} $ is less than $\sim$40\% and that for $\tk_{y,{\rm crit}}$ is less than  $\sim$50\%.

\section{Equations of the secular GI}
\label{appen}

We summarize the basic equations and the dispersion relation for a self-gravitating dust-gas disk to evaluate the most unstable wavelength of the secular GI.
We use the linearized basic equations derived by \cite{2019ApJ...881...53T}:
\begin{equation}
 n\delta \Sigma + ik_x\Sigma \delta u_{{\rm g},x} =0,
\end{equation}
\begin{equation}
 n\delta u_{{\rm g},x} = 2 \Omega \delta u_{{\rm g},y} 
-\frac{\cs^2}{\Sigma}ik_x\delta\Sigma -ik_x\delta \Phi_{\rm gd} 
-\frac{4}{3}\nu k_x^2\delta u_{{\rm g},x}
+\epsilon\frac{\delta u_{{\rm d},x}-\delta u_{{\rm g},x}}{\ts},
\end{equation}
\begin{equation}
 n\delta u_{{\rm g},y} = -\frac{\Omega}{2}\delta u_{{\rm g},x}-\nu k_x^2 \delta u_{{\rm g},y} -ik_x\frac{3\nu\Omega}{2\Sigma}\delta\Sigma+\epsilon\frac{\delta u_{{\rm d},y}-\delta u_{{\rm g},y}}{\ts},
\end{equation}
\begin{equation}
 n\delta\sd+ik_x\sd\delta u_{{\rm d},x} = -D k_x^2\delta\sd,
\end{equation}
\begin{equation}
 n\delta u_{{\rm d},x} = 2\Omega\delta u_{{\rm d},y}-\frac{\cd^2}{\sd} ik_x\delta\sd - ik_x\delta \Phi_{\rm gd} - \frac{\delta u_{{\rm d},x}-\delta u_{{\rm g},x}}{\ts},
\end{equation}
\begin{equation}
n\delta u_{{\rm d},y} = -\frac{\Omega}{2}
\left(\delta u_{{\rm d},x}-\frac{ik_xD}{\sd}\delta\sd\right)
 - \frac{\delta u_{{\rm d},y}-\delta u_{{\rm g},y}}{\ts}, 
\end{equation}
\begin{equation}
 \delta \Phi_{\rm gd} = -\frac{2 \pi G}{k_x}\left(\frac{\delta \Sigma}{1+k_xH}+\frac{\delta\sd}{1+k_xH_{\rm d}}\right),
\end{equation}
where $\Sigma$ and $\sd$ are the surface densities of gas and dust; $u_{{\rm g},x}$ and $u_{{\rm g},y}$ are the gas velocities in the $x$ and $y$ direction; $u_{{\rm d},x}$ and $u_{{\rm d},y}$ are the dust velocities in the $x$ and $y$ direction; $\Phi_{\rm gd}$ is the gravitational potential caused by gas and dust; $\delta \Sigma,\ \delta \sd,\ \delta u_{{\rm g},x},\ \delta u_{{\rm g},y},\ \delta u_{{\rm d},x},\ \delta u_{{\rm d},y},\ \delta\Phi_{\rm gd}$ are their perturbations; $n$ and $k_x$ are the growth rate and the wavenumber of the perturbations; $\Omega$ is the Kepler frequency; $\cs$ is the sound speed; $\nu$ is the coefficient of the kinematic viscosity; $\epsilon$ is the dust-to-gas mass ratio; $D$ is the dust diffusivity due to the gas turbulence; $\cd$ is the velocity dispersion of the dust; $G$ is the gravitational constant; and $H$ and $H_{\rm d}$ are the scale height of the gas and dust, respectively.
Here we assume that the perturbations are proportional to $\exp[nt+ik_xx]$.
When we neglect the back-reaction from dust to gas and use the approximations
$\alpha \ll 1$, St$\ll 1$, and $n/\Omega \gg 1$, we obtain the approximate dispersion relation given by Equation (\ref{eq:SGIapp}).

\bibliographystyle{aasjournal}
%\bibliography{ref}{}

\end{document}